\documentclass[prl,twocolumn,showpacs]{revtex4}
\usepackage{graphicx}
\begin{document}
\title{Giant infrared intensity of the Peierls mode  at the neutral-ionic
 phase transition}
 
\author{Luca \surname{Del Freo}}
%\altaffiliation{INSTM-UdR Parma}
\affiliation{Dip. di Chimica GIAF
 Universit\`{a} di Parma, I--43100 Parma, Italy; INSTM UdR Parma}
\author{Anna Painelli}
\email{anna.painelli@unipr.it}
%\altaffiliation{INSTM-UdR Parma}
\affiliation{Dip. di Chimica GIAF
 Universit\`{a} di Parma, I--43100 Parma, Italy; INSTM UdR Parma}
\author{Z.G.Soos}
\affiliation{ Dept. of Chemistry, Princeton University,
 Princeton, New Jersey 08544}
 
\date{\today}

 \begin{abstract}
 We present exact diagonalization results on a modified Peierls-Hubbard
 model
 for the neutral-ionic phase transition.  The ground
 state potential energy surface and the infrared intensity of the Peierls
 mode point to a strong, non-linear electron-phonon coupling,
 with effects that are dominated by the proximity to the electronic
 instability rather than by electronic correlations.
 The huge infrared intensity
 of the Peierls mode at the ferroelectric transition is related to
 the temperature dependence of the
 dielectric constant of mixed-stack organic crystals.
 \end{abstract}
 
 
\maketitle  
 
The nature of electron-phonon (e-ph) coupling in correlated
electron systems is an enduring topic in condensed matter research, 
currently as experimental indications of strong,  anharmonic e-ph coupling
have been recognized  in high-T$_c$ superconductors based on CuO  
\cite{lanzara} or the novel MgB$_2$ \cite{yildirim}. 
The topic itself is
 old and deeply rooted in the problem of structural deformation 
 in systems close to electronic instabilities, where particularly high 
sensitivity to e-ph coupling is expected.
 Low-dimensional systems have a host of electronic instabilities 
such as Mott, Peierls, and spin-Peierls transitions
  \cite{mazu1} with characteristic vibrational signatures \cite{bozio}. 
All these instabilities are accompanied by lattice relaxation 
 representing the extreme consequence of the non-linearity of
 electronic responses to perturbation by phonons.
 
The neutral-ionic transition (NIT) in a rigid 
donor-acceptor (DA) chain is the boundary between a paramagnet and a 
diamagnet \cite{cpl79}, analogous to a Mott metal-insulator
transition.
Such electronic instability is again accompanied by structural
relaxation. NITs were observed in organic charge-transfer (CT)
crystals with
mixed  DA stacks \cite{torrance}, and, as recognized from the outset
\cite{ttfca},  dimerization  accompanies the transition, leading to
ferroelectric (FE) behavior \cite{tokura}  for proper interstack arrangement. 
Recently NIT models were proposed for FE oxides \cite{egami}.
In this context the anomalous role of dimerization
in displacing electronic charge has been underlined, to suggest
enhanced e-ph coupling near the NIT. This was 
confirmed by subsequent  calculations \cite{rs} without, however,
pursuing experimental implications.

In this Letter, we exploit the recent definition \cite{resta,rs} 
of polarization, $P$, in insulators with periodic boundary conditions
(PBC) to obtain the infrared (IR) intensity of the dimerization mode.
The IR intensity depends sensitively on structure, which we relax
fully, and peaks at the FE transition of either correlated or uncorrelated
models, clearly implicating the structural instability rather than
correlations for the huge intensity. Working with a  \textit{linear} 
and \textit{harmonic} model for e-ph
coupling, we recognize large non-linearity  
in the anharmonicity of the ground state (GS) potential energy surface
(PES), and within the \textit{adiabatic} approximation we demonstrate an 
intimate entanglement of electrons and phonons. Sizeable non-adiabatic 
effects are expected near the NIT \cite{borghi}, but the subtle 
interplay between electrons and phonons we discuss here sets the baseline
for identifying non-adiabatic effects.

 We describe NIT in terms of a Peierls-Hubbard model with alternating
 on-site  energies ($\Delta$) and one electron per site. 
As in the Su-Schrieffer-Heeger (SSH) model \cite{ssh},
linear Peierls coupling to a $k=0$ phonon that
 modulates the transfer integrals is introduced in the following 
adiabatic Hamiltonian:
 
 \begin{eqnarray}
 {\mathcal H} &=& -\sum_{i,\sigma}\left(1+(-1)^i\delta
 \right)(c_{i,\sigma}^{\dagger}c_{i+1,\sigma}+h.c.)\label{ham}\\
 &+&U\sum_in_{i,\sigma}^{\dagger}n_{i,\sigma'}
 +\Delta\sum_{i,\sigma}(-1)^i c_{i,\sigma}^{\dagger}c_{i,\sigma}
 +\frac{N}{2\epsilon_d}\delta ^2 \nonumber
 \end{eqnarray}
 where $N$ is the number of sites, $c_{i,\sigma}^{\dagger}$
  creates   an electron with spin $\sigma$ on site $i$, and the last
 term accounts for the elastic energy.
 Accounting for alternating charges on molecular cores,
 $q_j$=2, 0 on D, A sites,
 respectively, we define the ionicity operator as
 $\hat\rho=\sum_j(q_j-\hat n_j)/N$.

 For $\Delta=0$ Eq.~(\ref{ham}) reduces to the
 SSH model \cite{ssh} in the $U =0$
 limit, or to a spin-Peierls model in the $U>>1$ limit, and
 the Peierls instability leads to finite dimerization amplitude, 
${\delta}$, in either case.
 The I  phase  with one
 electron per site ($\rho \sim 1$) is always dimerized 
\cite{nagaosa,paigir}.
 With increasing $\Delta$, electrons are paired
 on $-\Delta$ sites in the N phase and the GS is a regular chain
 (${ \delta} =0$).
 The equilibrium  $\delta$ decreases
 with $\Delta$ and vanishes for $\Delta$ larger than a critical
 value
 that depends on the lattice stiffness, as shown by  Rice and Mele for
 $U=0$ \cite{mr}.
 The correlated case is a Mott insulator on the I side \cite{anu},
 with an NIT to a band insulator around $\Delta \sim U/2$ in the regular
 chain.
Accordingly, we vary $\Gamma=\Delta -U/2$ in the vicinity of the NIT
 and solve ${\mathcal H}(\Gamma,\epsilon_d)$ exactly in the limit
 $\Delta, U \rightarrow \infty$ for finite chains with up to 12
sites  and PBC, as required for the $P$ operator \cite{resta}, using
the diagrammatic valence bond (VB) technique \cite{vb}. The lengthy
geometry optimization limits $N$, but is compensated by the rapid
convergence of GS properties compared to excitation energies \cite{anu}.

 In the rigid lattice ($\epsilon_d=0$)
$\rho$ varies  smoothly and   the  N-I crossover,  
at  $\Gamma_c=-0.668$ and $\rho \sim $ 0.63, is signaled by the 
unconditional dimerization instability of the I phase \cite{anu}.
\protect
\begin{figure}
\includegraphics [scale=0.38,angle=-90]{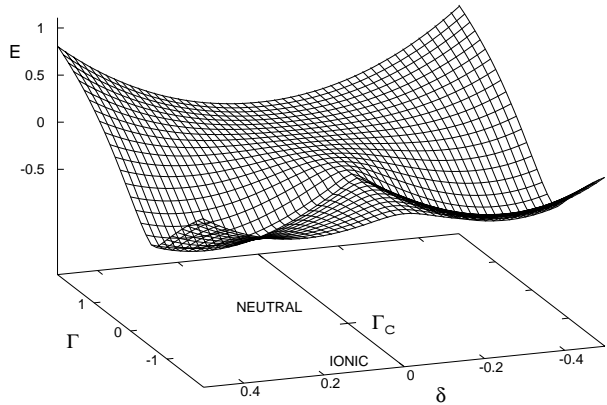}
\caption{ The GS energy of a 12 site ring with $\epsilon_d=0.64$.}
\label{fig1}
\end{figure}
\noindent
 A deformable lattice (finite $\epsilon_d$)
 dimerizes on the N side \cite{paigir}.
 Fig. \ref{fig1} depicts the evolution with $\Gamma$ of
 the GS PES of a 12-site ring with $\epsilon_d=0.64$.
 To focus attention on dimerization, the GS energy for the regular
 chain is set to zero at each $\Gamma$. Far in the N phase
 (large positive $\Gamma$) the
  regular stack is stable, but with decreasing  $\Gamma$,
 dimerization sets in well before  $\Gamma_c$, at
   $\rho \sim 0.25$. In the pre-transitional regime the GS PES
is clearly anharmonic.
 The two minima in the I regime  correspond to ferroelectrics 
 with opposite polarization.

To calculate the IR intensity of the Peierls mode, proportional 
to $(\partial P/\partial \delta)^2$,
we slightly modify the  Resta's twist operator  \cite{resta},  
to account for alternating charges on molecular cores, as follows:
 \begin{equation}
 P=\frac{1}{2\pi}Im (ln \langle\Psi
 |e^{i \frac{2\pi}{N}\hat{M}}|\Psi\rangle)
 =\frac{1}{2\pi}Im (ln (Z))
 \label{P}
 \end{equation}
 where $\Psi $ is the GS wavefunction, and
  $\hat{M}=\sum_j r_j (q_j-\hat n_j) $,  with $r_j$ locating the $j$
 site.
 The $\hat{M}$-operator is diagonal in the VB basis and
 the calculation of $P$ is easily implemented.
  $\hat{M}$ accounts for the actual geometry of the stack
 and depends on $\delta$ according to:
 \begin{equation}
 \hat{M}=\frac{N^2}{2}-\sum_j j \hat{n}_j
 +\frac{N\delta}{2\alpha}\hat{\rho}
 \label{m}
 \end{equation}
 where the mean intersite distance and the electronic charge are set to 1.
 The dimensionless e-ph coupling constant, $ \alpha= \sqrt{K\epsilon_d}$,
 is estimated $ \sim$ 5-10,  for typical values of the elastic constant, $K$.
 $\partial P/\partial \delta$ can be rigorously decomposed  into
 two contributions, one obtained by allowing $M$ to vary with $\delta$,
 at fixed $\Psi$, the other one by varying $\Psi$ at
 fixed $M$. The first contribution describes the IR intensity
 originating from the displacement of sites carrying a
 fixed charge, $\rho$. Frozen charges account for 
 IR intensity of optical modes in  ionic lattices.
 The second, electronic contribution instead accounts for the charge flux
 driven by $\delta$. It only appears in
 systems with mobile or delocalized electrons.

 For a regular chain, $\delta=0$,
 the reflection mirror residing on each site
 ensures real $Z$ values \cite{resta}.
 $Z$ is positive in the N regime, negative
 in the I regime, and $Z = 0$ locates the NIT of a $\delta=0$
 ring of any size with PBC \cite{rs}. The $\delta$ derivative
 of $P$ in Eq.~(\ref{P}) then diverges at the NIT, where the
 GS is metallic \cite{anu}. 
 In dimerized chains $Z$ is complex and the divergence of
 $\partial P/\partial \delta$ is suppressed. 
 Since the GS of Eq. (1) always dimerizes on the N side of the NIT, 
 the IR intensity remains finite for the equilibrium structure.
 The  IR intensity of the regular chain is then an upper bound 
 for the relaxed structure that, depending
 on $\epsilon_d$, holds on the N side of the FE transition. 

\begin{figure}
\includegraphics [scale=0.45,angle=0]{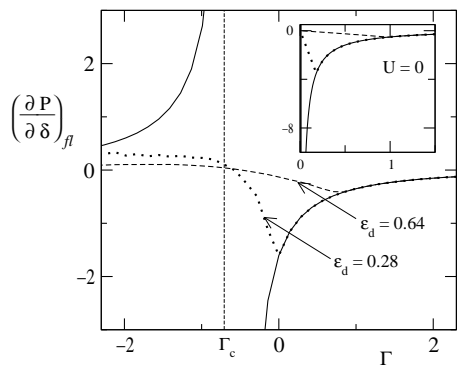}
 \caption{ The fluctuating charge  contribution to
 $\partial P/\partial \delta$, calculated
 for $N=$ 12 and  $U=\infty$.
 Continuous line refers to the regular chain. Dotted and dashed lines
 refer to the  equilibrium dimerization starting at the intersection
 with the continuous line. The inset shows results
 for $U=0$ and $N=\infty$.}
\label{fig2} 
\end{figure}

 We numerically evaluate  $\partial P/ \partial \delta$ at
 the equilibrium $ \delta$
 and  distinguish between contributions due to
 fluctuating and frozen charges. The latter is small 
 in the entire range of $\Gamma$. Large IR intensity is associated with
 fluctuating charges. Fig. \ref{fig2} shows  $ (\partial P/ \partial
 \delta)_{fl}$
 for 12-site chains. The $\delta=0$ curve 
 diverges
 at $\Gamma_c$ on both the N and I side. 
 \noindent
 Giant IR intensity at the NIT is due to finite electronic fluxes
 generated by infinitesimal
 lattice distortions.  Resta and Sorella reported similar results by
 finding the
 total average charge for fixed dimerization \cite{rs}; for small
 $\delta$,
 their quasi-divergent curves are finite-difference approximations to
 our  $\delta$-derivatives. 
  We see in Fig. \ref{fig2} that
 $ (\partial P/ \partial \delta)_{fl}=0$ near $\Gamma_c$ for
 both values of $\epsilon_d$.
 Dimerization increases the transfer integral $ (1 + \delta)$ and 
 tends to equalize charges on D and A sites then producing
 charge fluxes of opposite direction on the N and I side.
 $ (\partial P/ \partial \delta)_{fl}=0$
marks the point where charges are maximally delocalized and locates the
N-I crossover in dimerized chains.

 Since Mott insulators on the I side are inherently correlated, all
 NIT features have been viewed in terms of  correlations,
 including the divergence of dynamical charges
 at the NIT of the regular chain \cite{rs}.
 To address this point we investigated the non-interacting
 case, and found $P(\Delta,\delta,U=0)$ analytically for the
 infinite chain:
 \begin{equation}
 P( \Delta,\delta,U=0)
 =\frac{\delta \rho}{2\alpha} -\frac{4\delta}{\pi \Delta \sqrt{\Delta^2+4}}
 cel(q_c,p,a,b)
 \label{u0}
 \end{equation}
 where $cel(q_c,p,a,b)$ is the complete elliptic integral of the
 third kind \cite{recipes},
  calculated for $q_c^2=(\Delta^2+4\delta^2)/(\Delta^2+4)$,
 $p=(\Delta^2+4\delta^2)/(\Delta^2)$, $a=1$, and $b=0$.
 For the regular chain
 the $\delta$-derivative  of $\rho$ and of the elliptic
 integral above both vanish so that the frozen and
 the fluctuating charge contributions to
 $\partial P/\partial \delta$ are the coefficients of $\delta$ in
 the first and second term of Eq.~(\ref{u0}), respectively.
 For typical $\alpha$ values
 the frozen charge contribution smoothly increases from 0 to $\sim
 0.05$
 as $\rho$ increases from 0 to 1. The
 fluctuating charge contribution is more interesting and
  is shown  in the inset of Fig. \ref{fig2} for positive $\Delta$
($\Delta < 0$ simply reverses the role of D and A).  
$(\partial P/\partial\delta)_{fl}$
diverges as $\Delta \rightarrow$ 0, much as it does for $U=\infty$
 at NIT. Indeed, Eq.~(\ref{ham}) 
at $U=\Delta=0 $ describes the half-filled band
 whose instability Peierls pointed out, and $\Delta = 0 $ is the NIT
for free electrons. We conclude that divergent IR intensity at the FE
transition is not due to electronic correlations, 
but rather indicates a structural  instability that displaces 
electronic charges.
 The behavior of chains with finite $\epsilon_d$ at the equilibrium
dimerization
 (dashed and dotted lines in the inset) in the N phase is similar
 to the $U=\infty$ case, with
  $(\partial P/\partial \delta)_{fl}$ vanishing  at $\Delta=0$
and at $\Gamma_c$, respectively.

\begin{figure}
\includegraphics [scale=0.45,angle=0]{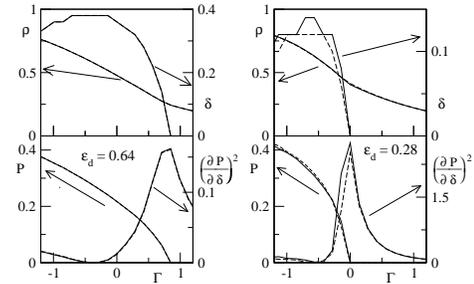}
  \caption{ GS equilibrium ionicity, dimerization amplitude, 
polarization
 and IR intensity  for $U=\infty$ and two different $\epsilon_d$.
 Note the
 scale change for both $\delta$ and $(\partial P/\partial \delta)^2$.
 Continuous and dashed lines refer to $N$=12 and 10, respectively (the two
 curves coincide in the left panels).}
\label{fig3} 
\end{figure}

 Fig. \ref{fig3} shows results for  Peierls coupling $\epsilon_d=$
0.28 and 0.64 in Eq.~(\ref{ham}) with $U=\infty$. 
The larger  dimerization for $\epsilon_d=$ 0.64
 strongly localizes electrons and reduces finite-size effects: rings
of 10 and 12 sites appear superimposed,  
while they show small differences
 for $\epsilon_d$ =0.28. The IR intensity is dominated by 
 fluctuating charges; it is small far in the N regime, 
peaks sharply at the FE transition,
 and  decreases rapidly to vanish at the N-I crossover.
% \protect
 \noindent
In fact, since the frozen charge
 contribution is  small,
 vanishing IR intensity can be taken as an operational definition
 of the charge crossover. 
The IR intensity smoothly increases again on the I side due
 to the
 increasing frozen charges and reaches typical values
 for ionic crystals as $\rho\rightarrow 1$.
 The peak in the IR intensity at the FE transition is larger the nearer
 the transition is to the NIT of the regular chain, where electronic
 charges are most responsive.
IR intensity does not measure the strength of e-ph coupling, 
nor the dimerization amplitude, rather it is governed by the response 
of the electronic system and reflects its non-linearity. 
 
  The Peierls-Hubbard model in Eq.~(\ref{ham}) contains neither intersite 
 electron-electron  interactions
 nor coupling to molecular (Holstein) modes.  Consequently 
 it cannot describe  materials such as TTF-CA \cite{ttfca} 
with a discontinuous NIT. But Eq.~(\ref{ham}) provides a useful 
starting point for systems with a continuous
 NIT, or, more precisely,
 a dimerization transition near a continuous NIT. 
Several CT salts dimerize in the N regime \cite{masino,dm,clbr3} with
 decreasing temperature or increasing pressure; this modulates
$\Gamma$  through the crystal's Madelung energy. 
Far IR data on these systems are however scanty,
the only available results having recently been
published on TTF-QBrCl$_3$ \cite{clbr3}.
This compound has $\rho\sim 0.3$ at ambient conditions and
 dimerizes at T$\sim$ 68 K. Concomitantly, its ionicity increases from
 $\sim$ 0.35 to 0.5 to reach $\sim$ 0.6 at T $\sim$ 10K.
Reflectivity measured at the lowest frequency (25 cm$^{-1}$)
 increases up to 90\% at the FE transition, suggesting
 the proximity to a metallic state there.
 After  dimerization, the reflectivity decreases again.
 Approaching the FE transition, a progressive displacement of the
 IR intensity towards lower frequencies can be inferred from
 data in Ref. \cite{clbr3}, and the published IR intensity
 vs. T closely resembles the  $\epsilon_d=$ 0.28 peak in Fig.
\ref{fig3}.
 Therefore we assign the amplified  far IR intensity  observed near the
FE transition to the soft dimerization mode, and explain its abrupt drop 
after the transition as due to the vanishing of the fluctuating
contribution 
to the IR intensity at the N-I crossover. This interpretation differs 
from the one proposed in Ref. \cite{clbr3}, where the far-IR 
absorption is assigned
to the dynamics of neutral-ionic domain-walls. Additional
experimental  and theoretical work is  needed to clarify this point.
 
 There are more data on static dielectric constants,
 $\varepsilon$ \cite{dm,clbr3}. Systems undergoing dimerization
 near a continuous NIT show a large peak in $\varepsilon$
 just where  long-range dimerization sets in.
 The similarity between the $\varepsilon$ peak (see e.g.
 Fig. 1 in ref. \cite{clbr3}) at the FE transition and the IR 
 intensity of the Peierls mode in Fig. 3 is suggestive.
E-ph coupling shifts oscillator strength from electronic states to
 lattice vibrations, and this displacement to lower
 energy amplifies the system's response to a static field.

 To  the first  order in $\epsilon_d$, the static response can be
 estimated as the sum of an electronic plus  a vibrational
contribution:
 $\varepsilon =1+(\chi_{el}+\chi_{vib})/\varepsilon_0$, 
with $\chi_{vib}=2\mu_{IR}^2/\omega \propto \epsilon_d (\partial P/
\partial \delta)^2$,
 defined in terms of the vibrational transition dipole moment and
 frequency \cite{cpl}.
 Analytical expressions of $\chi_{el}$ are available for $U=0$  \cite{caf}, and
 Fig. 4 compares the total (electronic+vibrational) $\varepsilon$
 with the purely electronic term. The lattice parameters, required to fix the 
absolute value of $\varepsilon$, are set to representative values for 
mixed stack crystals from Ref.~\cite{lecointe}.
 \protect
\begin{figure}
\includegraphics [scale=0.38,angle=0]{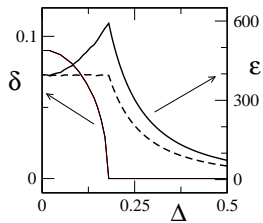}
\caption{ The equilibrium dimerization and  dielectric constant for $U=0$  
and $\epsilon_d=0.28$. Dashed and continuous line refer to the purely
electronic and to the total $\varepsilon$, respectively.
}
\label{fig4}
\end{figure}
\noindent
Whereas the magnitude of the two terms is comparable,
  the sharp peak at the FE transition is due e-ph coupling.
Again our interpretation differs from that in Ref. \cite{clbr3},
where the amplification of the dielectric constant at the FE transition was
related to the frequency of the far IR mode, via the Lyddane-Sachs-Teller
relation. We instead underline the role of the dimerization mode 
in displacing electronic charges.
  Consistent with available data \cite{dm,clbr3}, we predict large
 $\varepsilon$
 peaks at FE transitions occurring at intermediate $\rho$, i.e. for
 systems that, due to either  small Peierls coupling or to
 disorder, undergo dimerization very close to the
 electronic  (NIT) instability. 
 Preliminary $\varepsilon$ calculations on $U=\infty$ open chains yield
 similar results, suggesting once more that the physics of FE
 transitions  has more to do with phonons
 than with electronic correlations.

Experimental evidence is accumulating for strong
e-ph coupling in high-T$_c$ superconductors \cite{lanzara,yildirim}.
 The proximity
to an electronic instability is a unifying features of exotic
 superconductivity in cuprates, organics, or MgB$_2$, as well as
 IR intensity or dielectric responses of CT salts.
 Using an \textit{ adiabatic} model for an extended 
electronic system with an electronic
 instability, we demonstrate here that, quite irrespective of
electronic correlations, \textit{ linear}
e-ph coupling, \textit{harmonic} phonons and delocalized electrons
 produce large and non-linear effects
originating from a complex interplay of electronic and nuclear degrees
of freedom.

\begin{acknowledgments}
 We thank  Italian MIUR and CNR for supporting work in Parma;
 partial support from NSF through MRSEC program is acknowledged
 for work in Princeton.
\end{acknowledgments}


\begin{thebibliography}{25}
 \bibitem{lanzara} A.Lanzara \textit{ et al.}, 
  Nature, \textbf{ 412}, 510 (2001);
 R. J. McQueeney, \textit{ et al.}, Phys.~Rev.~Lett. \textbf{ 87},
 77001 (2001).
 \bibitem{yildirim} T.~Yildirim, \textit{ et al}, 
  Phys.~Rev.~Lett. \textbf{ 87},
 37001 (2001); 
 K.-P.Bohnen,  \textit{ et al}, Phys.~Rev.~Lett. \textbf{86},
 5771 (2001); 
 A.Y.Liu, \textit{ et al}, Phys.~Rev.~Lett. \textbf{87},
 87005 (2001)
\bibitem{mazu1} D.Baeriswyl, \textit{ et al}, in \textit{
Conducting Polymers}, edited by H.Kiess 
(Springer-Verlag, Heidelberg, 1992), p. 7%.

\bibitem{bozio} R.Bozio and C.Pecile, in
\textit{ Advances in Spectroscopy}, vol. 19 (wiley,
New York, 1991) p.1, and references therein.


\bibitem{cpl79} Z.G.Soos, \textit{ et al}, Chem.~Phys.~Lett., {\bf 65}, 
331 (1979).

 \bibitem{torrance} J.B. Torrance, \textit{ et al},  Phys.~Rev.~Lett.,
 \textbf{ 46}, 253
 (1981);  \textbf{ 47}, 1747 (1981).
 
 \bibitem{ttfca} A.Girlando, \textit{ et al}, J.~Chem.~Phys., \textbf{ 79}, 1075
 (1983).

\bibitem{tokura} S.~Horiuchi,   \textit{ et al}, J.~Am.~Chem.~Soc., 
{\bf 120}, 7379 (1998). 
 
 \bibitem{egami} T.~Egami,  \textit{ et al}, Science, {\bf 261}, 1307
 (1993).
 
 \bibitem{rs} R.Resta and S.Sorella,  Phys.~Rev.~Lett., \textbf{ 74}, 4738
 (1995).


 \bibitem{resta} R.Resta, Rev. Mod. Phys. \textbf{ 66}, 899 (1994).

\bibitem{borghi} G.P.Borghi,  \textit{ et al}, Europhys.~Lett., \textbf{ 34}, 
 127  (1996); 
 
 \bibitem{ssh} A.J.Heeger,  \textit{ et al}, Rev. Mod. Phys. \textbf{ 60}, 781
 (1988).

 
\bibitem{nagaosa} N.Nagaosa, J.Phys.Soc.Jpn, \textbf{ 55}, 2754 (1986);
\textbf{ 55} 3488 (1986).

 \bibitem{paigir} A.Painelli, and A.Girlando, Phys.~Rev.~B, \textbf{ 37},
 5748
 (1988).
 
 \bibitem{mr} M.J.Rice, and E.J.Mele,  Phys.~Rev.~Lett., \textbf{ 49}, 1455
 (1982).
 
 \bibitem{anu} Y.Anusooya-Pati,  \textit{ et al},  Phys.~Rev.~B,
 \textbf{ 63}, 205118 (2001); and references therein.
 
 \bibitem{vb} Z.G.Soos and S.Ramasesha, in \textit{ Valence Bond Theory and
 Chemical Structure}, edited by D.J.Klein and N.Trinajstic (Elsevier,
 New York, 1990), p. 81.
 
 \bibitem{recipes} W.H.Press, \textit{ et al.}, \textit{ Numerical Recipes},
 Cambridge University Press, Cambridge (1986).
 
 \bibitem{masino} A.Girlando, \textit{et al.},
 J.Chem.Phys. \textbf{ 98}, 7692 (1993);
  M.Masino \textit{et al.}, Phys.Chem.Chem.Phys.
 \textbf{ 3}, 1904 (2001); L.Farina, \textit{et al.},
 Phys. Rev. B \textbf{ 64}, 144102 (2001).
 
 \bibitem{dm} S. Horiuchi, \textit{et al.}, J.Am.Chem.Soc. \textbf{ 123}, 665 (2001).
 
 \bibitem{clbr3} Y.Okimoto,  \textit{ et al}, Phys.~Rev.~Lett., \textbf{ 87}, 187401 (2001).
 
 \bibitem{cpl} A.Painelli, Chem.Phys.Lett., \textbf{ 285}, 352 (1998)
 
 \bibitem{caf} C.Cojan,  \textit{et al.},
 Phys.Rev.B \textbf{ 15}, 909 (1977).

\bibitem{lecointe} M.~Le~Cointe,  \textit{et al.}, Phys. Rev. B 
\textbf{ 51}, 3374 (1995).
 
 \end{thebibliography}
 \end{document}